\documentclass[a4paper,12pt]{spieman}  % use this instead for A4 paper
\usepackage{amsmath,amsfonts,amssymb}
\usepackage{graphicx}
\usepackage{setspace}
\usepackage{tocloft}

\title{NeDF: neural deflection fields for sparse-view tomographic background oriented Schlieren}

\author[a]{Jiawei Li}
\author[b]{Xuhui Meng}
\author[a,c,*]{Yuan Xiong}
\author[a]{Tong Jia}
\author[a]{Chong Pan}
\author[a]{Jinjun Wang}
\affil[a]{Beihang University, Key Laboratory of Fluid Mechanics of Ministry of Education \& Aircraft and Propulsion Laboratory, Ningbo Institute of Technology, 37 Xueyuan Road, Beijing, China, 100191}
\affil[b]{Institute of Interdisciplinary Research for Mathematics and Applied Science, School of Mathematics and Statistics, Huazhong University of Science and Technology, 1037 Luoyu Road, Wuhan, China, 430074}
\affil[c]{Tianmushan Laboratory, 166 Shuanghongqiao Street, Pingyao Town, Hangzhou, China, 311115}

\cftpagenumbersoff{figure}
\cftpagenumbersoff{table} 
\begin{document} 
\maketitle

\begin{abstract}
Three-dimensional (3D) density-varying turbulent flows are widely encountered in high-speed aerodynamics, combustion, and heterogeneous mixing processes. Multi-camera-based tomographic background-oriented Schlieren (TBOS) has emerged as a powerful technique for revealing 3D flow density structures. However, dozens of cameras are typically required to obtain high-quality reconstructed density fields. Limited by the number of available optical windows and confined space in the harsh experimental environments, TBOS with only sparse views and limited viewing angles often becomes the necessary choice practically, rendering the inverse problem for TBOS reconstruction severely ill-posed and resulting in degraded tomography quality. In this study, we propose a novel TBOS reconstruction method, neural deflection field (NeDF), utilizing deep neural networks (DNNs) to represent the density gradient fields without using any pretrained neural network models. Particularly, state-of-the-art positional encoding techniques and hierarchical sampling strategies are incorporated to capture the density structures of high spatial frequencies. Required background images for TBOS reconstructions are synthesized based on a high-fidelity nonlinear ray-tracing method with the ground truth flows from conducting LES simulations on premixed turbulent flames. Owing to these synthesized BOS images, the superiority of the proposed method is quantitatively verified compared to the classical TBOS reconstruction methods, and the specific contributions from the position encoding and the hierarchical sampling strategy are also elucidated.
\end{abstract}

% Include a list of up to six keywords after the abstract

% Include email contact information for corresponding author
{\noindent \footnotesize\textbf{*}Yuan Xiong,  \linkable{xiongyuan@buaa.edu.cn} }

\begin{spacing}{2}   % use double spacing for rest of manuscript

\section{Introduction}
\label{sect:intro}  % \label{} allows reference to this section

Density-varying turbulent flows are widely encountered in various engineering fields, such as high-speed aerodynamics \cite{leon_three-dimensional_2022}, combustion \cite{grauer_instantaneous_2018}, and heterogeneous mixing \cite{livescu_variable-density_2008}. Exploring the spatial-temporal evolution of density fields proves essential for understanding the underlying flow physics. In fact, Schlieren-type methods, visualizing the light-of-sight density gradient, have played a crucial role in the history of high-speed aerodynamics \cite{settles_review_2017}. However, to avoid the line-of-sight-caused ambiguity in measuring turbulent flows, a non-intrusive optical technique is still highly desired to fetch the 3D density field quantitatively. For gaseous flows, flow density $\rho$ links to the local refractive index $n$ according to the Gladstone-Dale relation %as shown in (\ref{Eq:GD-relation})
\begin{equation}
n-1 =G\rho
\text{.}
\label{Eq:GD-relation}
\end{equation}
Inside the Gladstone-Dale constant $G$ depends on the light wavelength and the local gaseous composition. Thus, for gaseous flow, the density measurement can be converted to the measurement of the refractive index under certain conditions.

Schlieren and interferometry-based tomography techniques capable of reconstructing 3D refractive index fields have emerged for decades \cite{agrawal_three-dimensional_1998,leon_three-dimensional_2022}. However, similar attempts received only limited attention, possibly due to the complex optical setup involved. Around the year 2000, a new type of Schlieren method emerged known as synthetic Schlieren \cite{dalziel_whole-field_2000} or more popular as background-oriented Schlieren (BOS) technique \cite{richard_principle_2001, meier_computerized_2002}, Compared to the classical Schlieren methods, BOS significantly simplifies the experimental setup, optical alignment, and calibration procedures, thus allowing multi-camera based tomographic measurement with moderate setup complexity \cite{atcheson_time-resolved_2008}.

An essential step for multi-camera-based TBOS measurements involves discretizing the continuous $n$ field, usually cube-shaped, into discrete voxels, assuming either piecewise constant or radial basis function distributions inside. To achieve a high spatial resolution, the number of voxels(unknowns) can easily reach the level from $10^6$ to $10^9$ \cite{nicolas_direct_2016}. Considering that each camera view can typically provide about $10^5$ ray deflection vectors, in order to faithfully reconstruct multi-scaled turbulent density fields, dozens of camera views are thus required to alleviate the underdetermination of the tomographic reconstruction problem \cite{atcheson_time-resolved_2008, nicolas_direct_2016,li_three-dimensional_2024}. In addition, to maximize the effective information provided by each camera view and further alleviate the ill-posedness of the tomography, diversified locations of the camera views are encouraged compared to ones located only within a limited viewing angle \cite{nicolas_direct_2016,lang_measurement_2017}.

Extending TBOS measurements to density-varying flows like the ones within combustion engines and wind tunnels has gained increasing interests. However, limited by the number of available optical windows and confined working space in the harsh experimental environments, TBOS has to cope with only a few views within a limited viewing angle, resulting in a significantly degraded reconstruction quality due to the severe ill-posedness of the tomographic reconstruction. Instead of assuming complete independence between discrete voxels, a regularization approach could alleviate the ill-posedness of TBOS reconstruction by introducing certain prior knowledge of the flow, either via the visual hull technique to limit flow reconstruction volume \cite{atcheson_time-resolved_2008} or via regularization techniques according to CFD simulations \cite{grauer_instantaneous_2018} or empirical flow smoothness assumptions \cite{nicolas_direct_2016}.

DNNs have proven to be efficient in approximating high-dimensional functions. Using DNNs to represent the $n(\mathbf{x})$ fields, instead of allowing each voxel $n$ to vary independently, is akin to performing a reduced-order approximation on the $n$ fields, thus drastically reducing the number of unknowns to be reconstructed~\cite{lin_seamless_2021,karniadakis_physics-informed_2021,cui_enhancing_2024}. Ref.~\citenum{cai_flow_2021} pioneered in integrating TBOS and DNNs in the framework of physics-informed neural networks (PINNs). Velocity and pressure fields were inferred from a TBOS-reconstructed 3D density field via the PINNs. A classical conjugated gradient least square (CGLS) algorithm was utilized to reconstruct the density fields, together with the visual hull and a Tikhonov prior. Although the physically sound physical fields are inferred from PINN, the reconstructed temperature~(density) fields were still degraded due to the fact that only 6 camera views were utilized. Ref.~\citenum{molnar_estimating_2023} represented the density fields with DNNs in the framework of PINNs by exploring a simple axisymmetric supersonic flow. Only one BOS camera view was required, and the resulting deflection vector was used to build the loss function and train the PINN, from which all density, velocity, and pressure fields were inferred. While for sparse-view TBOS, measured light deflection vectors from a few camera views are insufficient to reconstruct a high-quality density field, the effectiveness of extracting more field variables on top of the density field requires further explorations, especially for complex multi-scale turbulent flows. 

In this paper, we will focus on the TBOS reconstruction of $\boldsymbol{\nabla} n$ field via DNNs. To cope with the TBOS measurement challenges due to sparse views and limited viewing angles, state-of-the-art DNNs strategies from the machine vision community, mainly based on NeRF \cite{mildenhall_nerf_2022}, are incorporated to build a brand-new neural deflection field (NeDF) model for TBOS. To quantitatively evaluate the performance of NeDF, LES simulated turbulent swirling flames serve as the ground truth flow fields. A series of high-fidelity synthetic BOS images are generated from different camera views, followed by a quantitative tomography error analysis that verified the superiority of the proposed NeDF. The paper is organized as follows:
the methodology is explained in Sec.~\ref{sec:Method}. The NeDF reconstructed results are presented and discussed in Sec.~\ref{sec:Results} while the conclusion is presented in Sec.~\ref{sec:Conclusion}. A supplementary material is also provided to explain the definition of evaluation metrics, training hyperparameter settings, and network architectures.

\section{Methodology}
\label{sec:Method}

\subsection{Sythetic BOS Image Generation}
\begin{table}
  \caption{Parameters used to synthesize the BOS images}
  \begin{center}
\def~{\hphantom{0}}
  \begin{tabular}{lllll}
      \hline
      \textbf{Parameter}  & \textbf{Setting}  &  & \textbf{Parameter}  & \textbf{Setting} \\[3pt]
      % \hline
      \cline{1-2} \cline{4-5}
      % \cmidrule(r){1-2}
       $L_{CB}$ &  $1520$ mm & & Sensor Res. & $560 \times 560$ pixels\\
       $L_{CO}$ & 850 mm & & Pixel Pitch & $11 \times 11$ $\mu$m$^2$\\
       Voxel Res.  & $161 \times 161 \times 159$ voxels &  &$\boldsymbol{\Delta}$ Res. & $560 \times 560$\\
       Volume Size   & $80 \times 80 \times 79$ mm$^3$ & & Cam. Config. & Coplanar Circular/Arc\\
       $f_{\mathrm{Lens}}$ & $50$ mm & & Cam. Num. & 5--180\\
       $f_{\#}$ & 32 & & Dot Num. & $100 \times 100$ \\
       Ray Num. & $2 \times 10^4$/dot & & Dot diamter & $360$ $\mu$m \\
       \hline
  \end{tabular}
  \label{tab:synthetic-TBOS}
  \end{center}
\end{table}

\begin{figure}
\centerline{\includegraphics[width=0.9\textwidth]{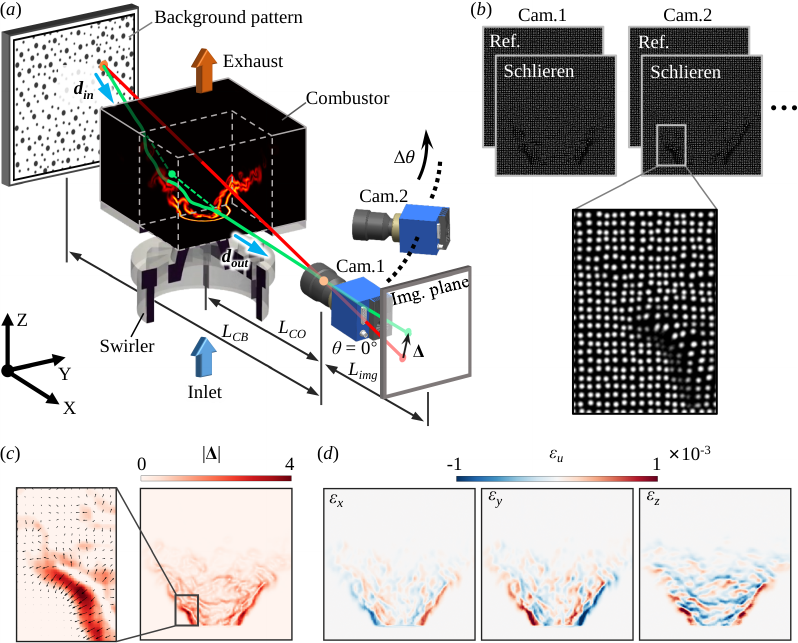}}
  % \captionsetup{justification=justified, singlelinecheck=false, width=\textwidth}
  \caption{schematic of (\textit{a}) TBOS principle \& Device setting; (\textit{b}) BOS images with/without flames from different views; (\textit{c}) Dot displacements on the sensor plane; (\textit{d}) Deflection vectors calculated based on the dot displacements.}
\label{Fig:Device_WorkFlow}
\end{figure}

To provide ground truth $\boldsymbol{\nabla} n$ fields for evaluating the NeDF performance quantitatively, large eddy simulations (LES) were conducted to simulate lean premixed turbulent swirl flames within the DLR aero-engine burner \cite{meier_detailed_2007} as detailed in Ref.~\citenum{li_three-dimensional_2024}. According to the BOS settings in Table~\ref{tab:synthetic-TBOS}, randomly dotted images with/without the swirling flames are synthesized as shown schematically in Fig.~\ref{Fig:Device_WorkFlow}(a) utilizing the nonlinear ray-tracing platform developed by Ref.~\citenum{rajendran_pivbos_2019}. Compared to the original code of Ref.~\citenum{rajendran_pivbos_2019}, we adopt a double-precision-based $3^{\text{rd}}$ order spline spatial interpolation together with the $4^{\text{th}}$ order Runge-Kutta scheme to solve the ray equation (\ref{Eq:RayEq}) \cite{atcheson_time-resolved_2008}, 
\begin{equation}
\frac{\displaystyle \mathrm{d}}%
     {\mathrm{d} s}
\left(n \frac{\displaystyle \mathrm{d} \boldsymbol{p}}%
     {\mathrm{d} s}\right) = 
\frac{\displaystyle \mathrm{d}}%
     {\mathrm{d} s}
\left(n \boldsymbol{d}\right) = \boldsymbol{\nabla} n \text{.}
\label{Eq:RayEq}
\end{equation}
Inside $\boldsymbol{p}$ is the position of a ‘particle’ traversing the light ray, $\mathrm{d} s$ is the differential ray path, $\boldsymbol{d}$ is the local ray direction. Equation~(\ref{Eq:RayEq}) can be further integrated to define the deflection vector $\boldsymbol{\varepsilon}$ as
\begin{equation}
\boldsymbol{\varepsilon} \approx \boldsymbol{d}_{\mathrm{out}} - \boldsymbol{d}_{\mathrm{in}} \approx \int_S \boldsymbol{\nabla} n \mathrm{d}s \text{,}
\label{Eq:RayDefl}
\end{equation}
where the subscript “$\mathrm{in}$” and “$\mathrm{out}$” indicate the entrance and exit positions of the ray on the reconstructed volume, and $n_{\mathrm{in}} \approx n_{\mathrm{out}} \approx 1$ is assumed for simplicity. According to Eq.~(\ref{Eq:RayDefl}), the ray deflection vector $\boldsymbol{\varepsilon}$ relies only on $\boldsymbol{\nabla} n$. Thus, we focus on reconstructing $\boldsymbol{\nabla} n$ rather than $n$ in this paper since the $n$ field can be computed straightforwardly by solving a 3D Poisson equation provided $\boldsymbol{\nabla} n$ with a reference $n$ value.

Synthesized BOS images with/without the flames are further processed using the wavelet-based optical flow analysis method proposed by Ref.~\citenum{schmidt_wavelet-based_2021} to compute the displacement vector $\boldsymbol{\Delta}$ (see Fig.~\ref{Fig:Device_WorkFlow}({c})), from which $\boldsymbol{\varepsilon}$ can be calculated straightforwardly based on $\boldsymbol{d}_{\mathrm{out}}$ and $\boldsymbol{d}_{\mathrm{in}}$ as shown schematically in Fig.~\ref{Fig:Device_WorkFlow}({a}). Further, by varying camera viewing angle $\theta$, $\boldsymbol{\varepsilon}$ from different views can be computed for subsequent tomography. The principle and procedures for classical TBOS reconstruction can be found in previous TBOS studies \cite{atcheson_time-resolved_2008,nicolas_direct_2016,li_three-dimensional_2024}. 

%which discretizes (\ref{Eq:RayDefl}) based on selected basis function $\Phi$ as
%\begin{equation}
%    \varepsilon_u
%    = \int_S \sum_{j=1}^J \left( \nabla n \right)_u^j \Phi^j \mathrm{d}s
%    = \sum_{j=1}^J \left[ \left( \nabla n \right)_u^j \int_S \Phi^j \mathrm{d}s \right] \text{,}
%\label{Eq:RayDefl_Discrete}
%\end{equation}
%where $u = \{x, y, z\}$, $J$ is the total number of discrete voxels, and $j$ is the voxel index. Considering a set of ray deflections $\boldsymbol{\varepsilon}_u \in \mathbb{R}^I$, (\ref{Eq:RayDefl_Discrete}) can be assembled into a linear system $\boldsymbol{\varepsilon}_u = \boldsymbol{T}  (\boldsymbol{\nabla}n)_u$, where $(\boldsymbol{\nabla}n)_u \in \mathbb{R}^J$ is the discrete field, and $\boldsymbol{T} \in \mathbb{R}^{I \times J}$ is the tomographic weight matrix. Obviously, the quality and efficiency of depend on the discretization precision and the choice of basis functions. If a step constant basis function is used to discretize the flow field used in the present work into uniform voxels ($0.5 \times 0.5 \times 0.5$ $\text{mm}^3/\text{voxel}$), the parameter space dimension $J$ exceeds $4 \times 10^6$, causing the linear equation system to be severely underdetermined during sparse BOS views. Moreover, the large scale of the tomographic matrix $\boldsymbol{T}$ also leads to storage difficulties and slow computation. Both the full circle camera arrangement ($\theta = 0^\circ \text{--} 360^\circ$) and arc arrangement ($\theta = 0^\circ \text{--} 155^\circ$) were tested to explore the impact from the limited viewing angle. 

\subsection{Neural Deflection Field}

\begin{figure}
  \centerline{\includegraphics[width=0.95\textwidth]{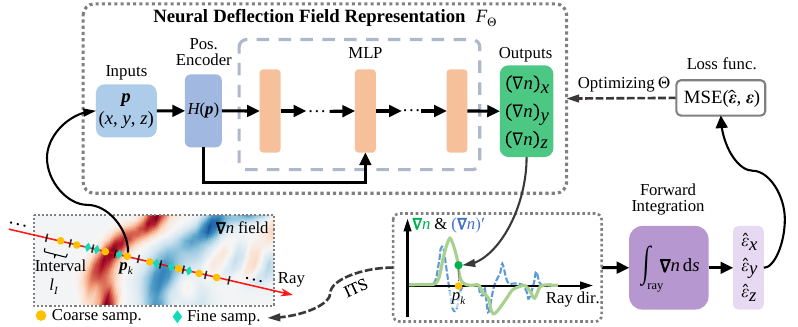}}
  % \captionsetup{justification=justified, singlelinecheck=false, width=\textwidth}
  \caption{Schematic of proposed NeDF model and training (tomographic reconstruction) workflow.}
\label{fig:NNmodel}
\end{figure}

Distinctive from classical TBOS reconstruction algorithms such as CGLS, simultaneous algebraic reconstruction technique (SART) and simultaneous iterative reconstruction technique (SIRT), we construct an implicit function based on the multilayer perceptron (MLP) for representing $\boldsymbol{\nabla} n$ as shown in Fig.~\ref{fig:NNmodel}. The mapping function $F_{\Theta}(\boldsymbol{p}) = (\nabla n)_u$ defines the neural deflection field (NeDF), where $\Theta$ represents the trainable parameters of NeDF, and inside $u=\{x,y,z \}$ represents coordinate directions. Taking one ray as an example, the training procedures of NeDF are briefly summarized as follows: (\textit{a}) Sample multiple positions $\{\boldsymbol{p}_1, \dots, \boldsymbol{p}_K\}$ along the ray; (\textit{b}) Predict the $\boldsymbol{\nabla} n$ through NeDF at the corresponding positions from (a); (\textit{c}) Calculate deflection vector ${\boldsymbol{\varepsilon}}$ using a forward integrator; (\textit{d}) Compute the mean squared error (MSE) between predicted $\hat{\boldsymbol{\varepsilon}}$ and observed ${\boldsymbol{\varepsilon}}$ as the loss function:
\begin{equation}
    \text{MSE}(\hat{\boldsymbol{\varepsilon}},\boldsymbol{\varepsilon})= \frac{1}{N}\sum_{i=1}^N (\hat{\varepsilon}_i - \varepsilon_i)^2 \text{,}
\end{equation}
where $N$ represents data size in a batch; and (\textit{e}) Estimate NeDF parameters $\Theta$ using the stochastic gradient descent (SGD) method. Inside, the neural forward projection (\textit{a})--(\textit{c}) distinguishes current NeDF from existing DNNs-based TBOS approaches, thus requiring further elaboration in the following.

The first highlighted feature of the NeDF relies on the abortion of spatially uniform sampling strategies. With the classical uniform voxel discretization, the targeted density field is discretized and sampled uniformly in space~\cite{nicolas_direct_2016}. Considering the turbulent premixed flames explored here, large density gradients concentrate only near the flame fronts, the thickness of which spans only less than $1$ mm. Uniform sampling of the entire density gradient fields inevitably reduces reconstruction quality by smoothing out the sharp transitions in $\boldsymbol{\nabla} \rho$ near the flame fronts while also resulting in unnecessary sampling in most regions with only trivial $\boldsymbol{\nabla} \rho$. Similar flow scenarios also exist for compressible shear layers and shock waves. In this regard, NeDF incorporates the hierarchical sampling strategy akin to Ref.~\citenum{mildenhall_nerf_2022} with detailed implementations as follows: (\textit{a}) divide the ray into $K_c$ intervals and sample once at random location inside to retain a large sampling spatial coverage with certain randomness. Based on this sampling, a network termed as "coarse network" $F_{\Theta_C}$ is trained, from which $(\nabla n)_u$ can be obtained by mapping $F_{\Theta_C} \{\boldsymbol{p}_1, \cdots ,\boldsymbol{p}_{K_C}\}$; (\textit{b}) define $\| (\nabla n)_u ' \| / \sum\|(\nabla n)_u '\|$ as the sampling probability density function and refine $K_F$ sampling points via inverse transform sampling (ITS) along the entire ray; (\textit{c}) combine both the coarse and refined sampling sets $\{\boldsymbol{p}_1, \cdots ,\boldsymbol{p}_{K_C + K_F}\}$ to jointly optimize $\{ F_{\Theta_C}, F_{\Theta_F} \}$ to achieve a refined $F_{\Theta_F}$. For the turbulent flames explored here, $K_C = 128$ ($l_I \approx 1$ mm) and $K_F = 64$ are tested to be sufficient.

The second highlighted feature of the NeDF is attributed to its effectiveness in retaining high-frequency spatial structures of the flow. As pointed out by Ref.~\citenum{rahaman_spectral_2019}, MLPs are inclined to learn the low-frequency features of the data. Complicating the data manifold in the network by mapping the network input to a higher-dimensional space could improve the MLP performance in the high-frequency range. In this regard, NeDF incorporates the multiresolution hash encoding as suggested by Ref.~\citenum{muller_instant_2022} to elevate the dimensionality of position coordinates before feeding them into the MLP. The encoding strategy segments coordinate space into multi-scale grids ($M$ levels), with vertices corresponding to a $F$-dimensional feature vector determined by a learnable hash table and spatial hash functions. Specifically, the positional encoding is formulated as
\begin{equation}
    H_\Omega(\boldsymbol{p}) = \left[ L\left(T_1(\boldsymbol{p})\right), L\left(T_2(\boldsymbol{p})\right), \cdots , L\left(T_M(\boldsymbol{p})\right) \right]^T \text{,}
\label{Eq:HashEncode}
\end{equation}
where $T(\boldsymbol{p})$ refers to the feature vectors of all vertices in the grids containing point $\boldsymbol{p}$, $L$ is the linear interpolation operator to obtain the $F$--dimensional vector for point $\boldsymbol{p}$, $\Omega$ is the encoding parameter. Via the encoding, 3D input $\boldsymbol{p}$ is transformed into an $M \times F$--dimensional vector. $M = 16$ and $F = 2$ are set here resulting in a $32$--dimensional encoding vector.

The implementation of NeDF is based on the PyTorch framework on a workstation equipped with an Intel Xeon w7-2475X CPU and an NVIDIA GeForce RTX 4090 GPU. The MLP consist of $5$ layers with $44$ nodes each, and the LeakyReLU activation function is applied in the hidden layers. Similar to Ref.~\citenum{mildenhall_nerf_2022}, the encoding vector is concatenated with the input of the $3^{rd}$ hidden layer. The network has $8935$ trainable parameters, initialized with the Kaiming uniform method. To improve the training accuracy, $\nabla n$ is normalized with $| \nabla n |_{max}$. A batch of $2048$ random rays from a view is selected to update the model, and a loop through all camera views is treated as one epoch. The Adam algorithm is utilized to minimize the loss function \cite{kingma_adam_2017}. The learning rate $l_r$ is set as a piecewise constant and decreases progressively with training, and the detailed $l_r$ schedule is documented in the supplementary material. The Softsign function is imposed at the output of the NeDF to guarantee $\Tilde{\nabla} n$ ranged from $-1$ to $1$. We have conducted a series of comparative studies on the hyperparameters (e.g., architectures of NeDF, learning rate, etc.) of NeDF to demonstrate that stable reconstruction results can be obtained using the present model (see details in the supplementary material). 

\section{Results and discussion}\label{sec:Results}

In this section, $\nabla n$ fields reconstructed by the NeDF are systematically compared to the ones obtained by classical TBOS reconstruction methods, including SIRT, CGLS, and CGLS with visual hull applied (CGLS-VH) at the same voxel locations. Details for implementing the classical algorithms and algorithm parameters can be found in our recent study \cite{li_three-dimensional_2024}, while the stop criterion for training the loss function of NeDF is set to $5 \times 10^{-9}$. To enable quantitative evaluation of the reconstruction quality, three metrics are used, including the root mean square error (RMSE) indicating the overall deviations from the ground truth, the peak signal-to-noise ratio (PSNR) characterizing the strength of the signal compared to the reconstruction noises, and the structural similarity index (SSIM) measuring the similarity between the two fields. Detailed definitions and parameters for the three metrics are in the supplementary material.

\begin{figure}
  \centerline{\includegraphics[width = 0.9\textwidth]{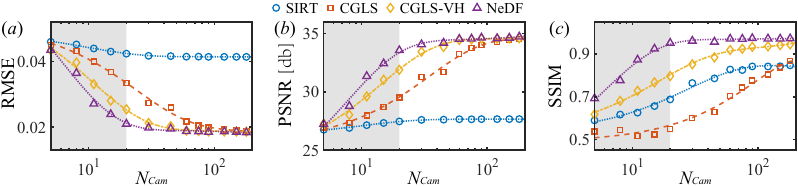}}
  % \captionsetup{justification=justified, singlelinecheck=false, width=\textwidth}
  \caption{Statistics of the reconstruction results ($\| \nabla n \|$) under the conditions of $5$--$180$ cameras and circular array: (\textit{a}) RMSE, (\textit{b}) PSNR, and (\textit{c}) SSIM. The dashed lines are the trend of exponential fitting.}
\label{Fig:MultiAngles_evaluations}
\end{figure}

To shed light on the performance of NeDF, especially its capability to cope with sparse views, an evenly distributed, circular-shaped, coplanar camera configuration is adopted by varying the camera number $N_{\mathrm{Cam}}$ from 5 to 180. The lower threshold $N_{\mathrm{Cam}} = 5$ represents an extreme sparse-view configuration since most TBOS studies for similar flows have set $N_{\mathrm{Cam}} \geq 9$. Figure~\ref{Fig:MultiAngles_evaluations} summarizes the dependency of the three evaluation metrics over $N_{\mathrm{Cam}}$. The reasonableness of the dependencies can be verified by identifying that CGLS-VH outperforms CGLS and SIRT owing to the VH prior. The effectiveness of VH in improving the reconstruction quality is widely acknowledged in TBOS studies \cite{atcheson_time-resolved_2008,nicolas_direct_2016,li_three-dimensional_2024} since the number of unknown voxels is effectively reduced by applying the VH constrain. 

The superiority of the proposed NeDF can thus be faithfully confirmed since its reconstructions give the lowest RMSE while retaining the highest PSNR and SSIM through all tested $N_{\mathrm{Cam}}$. Moreover, a unique region can be further identified as marked out by the shadow in Fig.~\ref{Fig:MultiAngles_evaluations}. For $N_{\mathrm{Cam}} \leq 20$, three metrics all exhibit linear dependencies on $10^{N_{\mathrm{Cam}}}$ demonstrating the extreme sensitivity of the tomography quality over $N_{\mathrm{Cam}}$. Thus, we can define $N_{\mathrm{Cam}} \leq 20$ as the `sparse-view' scenario for TBOS with the turbulent flames explored. In this scenario, as the number of camera views in the TBOS experiment decreases, the reconstruction errors increase more significantly with each further reduction in the number of views. Owing to the remarkable performance of NeDF, the RMSE and PSNR using NeDF with $N_{\mathrm{Cam}}=11$ are comparable to those obtained from CGLS-VH with $N_{\mathrm{Cam}}=18$, saving the view number by $39\%$. Similarly, NeDF with $N_{\mathrm{Cam}}=11$ can achieve SSIM comparable to that of CGLS-VH with $N_{\mathrm{Cam}}=35$, reducing the number of views by $69\%$.

\begin{figure}
  \centerline{\includegraphics[width=0.9\textwidth]{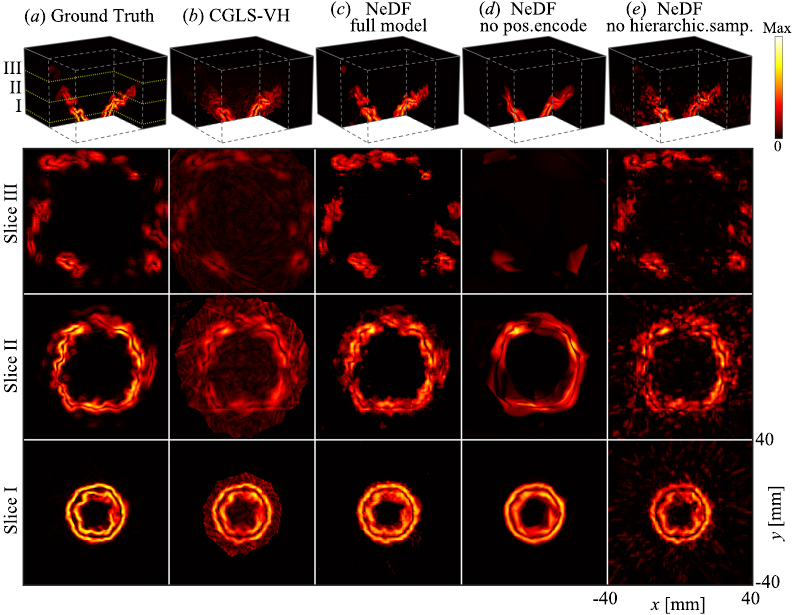}}
  % \captionsetup{justification=justified, singlelinecheck=false, width=\textwidth}
  \caption{Reconstruction results ($\|\nabla n\|$) under the condition of sparse-view and a limited viewing angle: the first row is the 3D data, and the rows $2$--$4$ correspond to the XY--slices at Z $=37$, $21$, and $5$ mm, respectively; The first column is ground truth, and columns $2$--$5$ correspond to the results of different reconstruction methods. Upper limits of dynamic range of rows $1$--$4$ are $\text{Max} = 0.8$, $0.4$, $0.6$ and $0.8$, respectively.}
\label{Fig:SparseViewResult_Slices}
\end{figure}

After clarifying the advantage of NeDF coping with sparse views, we further explore the performance of NeDF with a limited viewing angle for TBOS applications within confined spaces. An evenly distributed arc-shaped coplanar camera configuration is utilized with $N_{\mathrm{Cam}}=11$ and $\theta \in(0^{\circ}-155^{\circ})$. The $\|\nabla n\|$ fields from the ground truth LES simulation and the ones reconstructed by CGLS-VH and NeDF are listed in column (a)--(c) of Fig.~\ref{Fig:SparseViewResult_Slices} at three vertical locations. Overall, NeDF successfully captures both the large coherent flow structure as well as the local tiny wrinkles near the flame fronts. Comparing $\|\nabla n\|$ distributions in slice III downstream of the combustor, the excellent performance of NeDF in capturing trivial-small-scale density gradients can be verified compared to the CGLS-VH method. Note that obvious spike-like reconstruction noises from CGLS-VH can be further suppressed by carefully tuning the size of VH according to the empirical trial-and-errors, which is not implemented here. Comparisons between the three slices obtained by CGLS-VH and NeDF reveal another remarkable feature of NeDF: no need to establish a good-quality VH with repetitive trials according to the user's experiences.

To gain an insight into the remarkable performance of NeDF, the contributions from the position encoding and hierarchical sampling strategy are explored individually. The column (d)--(e) within Fig.~\ref{Fig:SparseViewResult_Slices} shows the reconstruction results from incomplete NeDF with position encoding and hierarchical sampling removed, respectively. By comparing the NeDF without position encoding with the complete NeDF model, the over-smoothed $\|\nabla n\|$ field can be immediately identified, verifying the effectiveness of position encoding in retaining the high-frequency spatial structures as discussed in  Sec.~\ref{sec:Method}. Interestingly, since the spike-like noises within CGLS-VH are mostly of high spatial frequency, these noises are filtered completely by removing position encoding, too. Moreover, by comparing the complete NeDF model to the one without hierarchical sampling, implemented by only training the coarse network $F_{\Theta_C}$, the spike-like spurious noises can be observed again owing to the position encoding. In addition, without the hierarchical sampling, attenuated peak values of $\|\nabla n\|$ can also be identified, especially in slice I, confirming the effectiveness of the hierarchical sampling in retaining the sharp gradients. 

\section{Conclusion}\label{sec:Conclusion}

To cope with the challenges of sparse views and limited viewing angles with multi-camera-based TBOS, a novel neural deflection field~(NeDF) was proposed without the need for pretrained DNNs. Within the NeDF, a deep neural network was utilized to approximate the high-dimensional $\nabla n$ fields of the reduced-order modelling fashion. This is in contrast to the classical TBOS, which allows discrete voxels of $\nabla n$ to vary independently. To quantitatively evaluate the performance of the proposed NeDF, we synthesized the BOS images based on a high-fidelity nonlinear ray-tracing method with ground truth flows provided from LES simulations on premixed turbulent flames within the DLR aero-engine combustor. Owing to the synthesized BOS images, we quantitatively verified the superiority of NeDF compared to the classical tomographic algorithms. The specific contributions from the position encoding and the hierarchical sampling strategy were also elucidated in NeDF, such that the position encoding manages to retain the high-frequency spatial structures, and the hierarchical sampling strategy retains the sharp peaks within the flow structures. In addition, the definition of `sparse-view' was proposed based on the dependency of three metrics over the camera view number, under the condition of which the TBOS reconstruction quality exhibits a high sensitivity over the number of camera views.

\subsection* {Acknowledgments}
We thank Prof. Wang Han and Prof. Yihao Tang from the School of Astronautics of Beihang University for providing LES data for the present study. 

\subsection* {Funding}
The present study was supported by the National Natural Science Foundation of China (NSFC Grant No. 12102028).

%%%%% References %%%%%

\bibliography{NeDF}   % bibliography data in report.bib

\begin{thebibliography}{10}

\bibitem{leon_three-dimensional_2022}
O.~Léon, D.~Donjat, F.~Olchewsky, {\em et~al.}, ``Three-dimensional density field of a screeching under-expanded jet in helical mode using multi-view digital holographic interferometry,'' {\em Journal of Fluid Mechanics} {\bf 947}, A36  (2022).

\bibitem{grauer_instantaneous_2018}
S.~J. Grauer, A.~Unterberger, A.~Rittler, {\em et~al.}, ``Instantaneous {3D} flame imaging by background-oriented schlieren tomography,'' {\em Combustion and Flame} {\bf 196}, 284--299  (2018).

\bibitem{livescu_variable-density_2008}
D.~Livescu and J.~R. Ristorcelli, ``Variable-density mixing in buoyancy-driven turbulence,'' {\em Journal of Fluid Mechanics} {\bf 605}, 145--180  (2008).

\bibitem{settles_review_2017}
G.~S. Settles and M.~J. Hargather, ``A review of recent developments in schlieren and shadowgraph techniques,'' {\em Measurement Science and Technology} {\bf 28}(4)  (2017).

\bibitem{agrawal_three-dimensional_1998}
A.~K. Agrawal, N.~K. Butuk, S.~R. Gollahalli, {\em et~al.}, ``Three-dimensional rainbow schlieren tomography of a temperature field in gas flows,'' {\em Applied Optics} {\bf 37}, 479  (1998).

\bibitem{dalziel_whole-field_2000}
S.~B. Dalziel, G.~O. Hughes, and B.~R. Sutherland, ``Whole-field density measurements by 'synthetic schlieren','' {\em Experiments in Fluids} {\bf 28}(4), 322--335  (2000).

\bibitem{richard_principle_2001}
H.~Richard and M.~Raffel, ``Principle and applications of the background oriented schlieren ({BOS}) method,'' {\em Measurement Science and Technology} {\bf 12}(9), 1576--1585  (2001).

\bibitem{meier_computerized_2002}
G.~Meier, ``Computerized background-oriented schlieren,'' {\em Experiments in Fluids} {\bf 33}(1), 181--187  (2002).

\bibitem{atcheson_time-resolved_2008}
B.~Atcheson, I.~Ihrke, W.~Heidrich, {\em et~al.}, ``Time-resolved 3d capture of non-stationary gas flows,'' {\em ACM Transactions on Graphics} {\bf 27}(5), 132:1--132:9  (2008).

\bibitem{nicolas_direct_2016}
F.~Nicolas, V.~Todoroff, A.~Plyer, {\em et~al.}, ``A direct approach for instantaneous {3D} density field reconstruction from background-oriented schlieren ({BOS}) measurements,'' {\em Experiments in Fluids} {\bf 57}, 13  (2016).

\bibitem{li_three-dimensional_2024}
J.~Li, Y.~Xiong, Y.~Tang, {\em et~al.}, ``Three-dimensional diagnosis of lean premixed turbulent swirl flames using tomographic background oriented {Schlieren},'' {\em Physics of Fluids} {\bf 36}, 055159  (2024).

\bibitem{lang_measurement_2017}
H.~M. Lang, K.~Oberleithner, C.~O. Paschereit, {\em et~al.}, ``Measurement of the fluctuating temperature field in a heated swirling jet with {BOS} tomography,'' {\em Experiments in Fluids} {\bf 58}, 88  (2017).

\bibitem{lin_seamless_2021}
C.~Lin, M.~Maxey, Z.~Li, {\em et~al.}, ``A {Seamless} {Multiscale} {Operator} {Neural} {Network} for {Inferring} {Bubble} {Dynamics},'' {\em Journal of Fluid Mechanics} {\bf 929}, A18  (2021).

\bibitem{karniadakis_physics-informed_2021}
G.~E. Karniadakis, I.~G. Kevrekidis, L.~Lu, {\em et~al.}, ``Physics-informed machine learning,'' {\em Nature Reviews Physics} {\bf 3}, 422--440  (2021).

\bibitem{cui_enhancing_2024}
X.~Cui, B.~Sun, Y.~Zhu, {\em et~al.}, ``Enhancing efficiency and propulsion in bio-mimetic robotic fish through end-to-end deep reinforcement learning,'' {\em Physics of Fluids} {\bf 36}, 031910  (2024).

\bibitem{cai_flow_2021}
S.~Cai, Z.~Wang, F.~Fuest, {\em et~al.}, ``Flow over an espresso cup: inferring 3-{D} velocity and pressure fields from tomographic background oriented {Schlieren} via physics-informed neural networks,'' {\em Journal of Fluid Mechanics} {\bf 915}, A102  (2021).

\bibitem{molnar_estimating_2023}
J.~P. Molnar, L.~Venkatakrishnan, B.~E. Schmidt, {\em et~al.}, ``Estimating density, velocity, and pressure fields in supersonic flows using physics-informed {BOS},'' {\em Experiments in Fluids} {\bf 64}, 14  (2023).

\bibitem{mildenhall_nerf_2022}
B.~Mildenhall, P.~P. Srinivasan, M.~Tancik, {\em et~al.}, ``{NeRF}: representing scenes as neural radiance fields for view synthesis,'' {\em Communications of the ACM} {\bf 65}, 99--106  (2022).

\bibitem{meier_detailed_2007}
W.~Meier, P.~Weigand, X.~R. Duan, {\em et~al.}, ``Detailed characterization of the dynamics of thermoacoustic pulsations in a lean premixed swirl flame,'' {\em Combustion and Flame} {\bf 150}, 2--26  (2007).

\bibitem{rajendran_pivbos_2019}
L.~K. Rajendran, S.~P.~M. Bane, and P.~P. Vlachos, ``{PIV}/{BOS} synthetic image generation in variable density environments for error analysis and experiment design,'' {\em Measurement Science and Technology} {\bf 30}(8), 085302  (2019).

\bibitem{schmidt_wavelet-based_2021}
B.~E. Schmidt and M.~R. Woike, ``Wavelet-{Based} {Optical} {Flow} {Analysis} for {Background}-{Oriented} {Schlieren} {Image} {Processing},'' {\em AIAA Journal} {\bf 59}, 3209--3216  (2021).

\bibitem{rahaman_spectral_2019}
N.~Rahaman, A.~Baratin, D.~Arpit, {\em et~al.}, ``On the {Spectral} {Bias} of {Neural} {Networks},'' in {\em Proceedings of the 36th {International} {Conference} on {Machine} {Learning}},  5301--5310, PMLR  (2019).

\bibitem{muller_instant_2022}
T.~Müller, A.~Evans, C.~Schied, {\em et~al.}, ``Instant {Neural} {Graphics} {Primitives} with a {Multiresolution} {Hash} {Encoding},'' {\em ACM Transactions on Graphics} {\bf 41}, 1--15  (2022).

\bibitem{kingma_adam_2017}
D.~P. Kingma and J.~Ba, ``Adam: {A} {Method} for {Stochastic} {Optimization},''  (2017).
\newblock arXiv:1412.6980.

\end{thebibliography}
\bibliographystyle{spiejour}   % makes bibtex use spiejour.bst

%%%%% Biographies of authors %%%%%

% \vspace{2ex}\noindent\textbf{First Author} is an assistant professor at the University of Optical Engineering. He received his BS and MS degrees in physics from the University of Optics in 1985 and 1987, respectively, and his PhD degree in optics from the Institute of Technology in 1991.  He is the author of more than 50 journal papers and has written three book chapters. His current research interests include optical interconnects, holography, and optoelectronic systems. He is a member of SPIE.

% \vspace{1ex}
% \noindent Biographies and photographs of the other authors are not available.

% \listoffigures
% \listoftables

\end{spacing}
\end{document}

\maketitle

% \begin{abstract}
% Your abstract.
% \end{abstract}

\section{Evaluation metrics}
\label{sec:metrics}

The mean squared error (MSE) between the  NeDF predictions $ x_i$ and experimental measurements $y_i$ is utilized as the loss function in the present study:
\begin{equation}
    \text{MSE}(x,y)= \frac{1}{N}\sum_{i=1}^N (x_i - y_i)^2 \text{,}
\end{equation}
where $N$ is the total number of data points in the array, and $i$ is the data index.

We use the root mean squared error (RMSE) to evaluate the overall deviation between the reconstructed data and the reference data:
\begin{equation}
    \text{RMSE}(x,y) = \sqrt{\text{MSE}(x,y)} \text{.}
\end{equation}

The peak signal-to-noise ratio (PSNR) is used to characterize the ratio between the power of the reconstructed noise and the maximum possible power of the signal, where $D$ is the dynamic range of the data:
\begin{equation}
    \text{PSNR}(x,y) = 10 \log_{10} \frac{D^2}{\text{MSE}(x,y)} \text{.}
\end{equation}

The structural similarity index (SSIM) is used to comprehensively measure the similarity between the data's amplitude, contrast, and structure:
\begin{equation}
    \text{SSIM}(x,y) = \frac{2\mu_x \mu_y + C_1}{\mu_x^2 + \mu_y^2 + C_1}
                       \cdot
                       \frac{2\sigma_x\sigma_y+C_2}{\sigma_x^2+\sigma_y^2 + C_2}
                       \cdot
                       \frac{\sigma_{xy} + C_3}{\sigma_x\sigma_x+C_3} \text{,}
\label{Eq:SSIM}
\end{equation}
where $\mu_x$ and $\mu_y$ are the means, and $\sigma_x$, $\sigma_y$, and $\sigma_{xy}$ are the standard deviations and cross-covariance of $x$ and $y$. $C_1$, $C_2$, and $C_3$ are regularization constants to prevent instability in conditions with means or standard deviations approaching zero. The first term on the right-hand side of Eq.(\ref{Eq:SSIM}) assesses the data amplitude, the second evaluates contrast, and the third assesses structural similarity. In this study, $C_1=(0.01D)^2$, $C_2=(0.03D)^2$, and $C_3=C_2/2$, and SSIM can be simplified as:
\begin{equation}
    \text{SSIM}(x,y) = 
        \frac{(2\mu_x \mu_y + C_1)(\sigma_{xy} + C_2)}
             {(\mu_x^2 + \mu_y^2 + C_1)(\sigma_x^2+\sigma_y^2 + C_2)} \text{.}
\end{equation}

\section{Learning rate schedule in NeDF}
\label{sec:lr}
The learning rate schedule, e.g., learning rate decay, is widely used in the deep learning community to improve the predicted accuracy of DNNs. However, we note that the optimal learning rate schedule which results in the best predicted accuracy of DNNs is problem-dependent and is generally difficult to obtain. We found empirically that the following learning rate ($l_r$) schedule in NeDFs is capable of providing accurate and robust results, which is expressed as follows:
\begin{equation}
    l_r = l_r^* \cdot \alpha^c, ~ (c-1) \cdot E_d + 1 \leq e \leq c \cdot E_d \text{,}
\end{equation}
where, $e \in [1,E]$ represents the training epoch,  $c= \lceil e/E_d \rceil$ is the segment count,  $l_r^*$ is the initial learning rate, $E$ denotes the total number of epochs, $E_d$ is the number of epochs per segment, and $\alpha$ is the learning rate decay factor.  In this study, we tested various cases with different numbers of cameras (views), ranging from $5$ to $180$. The hyperparameters, e.g., initial learning rate, etc., employed in different cases are summarized in Table \ref{Tab:Hyperparameters}.
    \begin{table}
    \centering
    \begin{tabular}{c|c|c|c|c|c|c|c|c|c|c|c|c|c}
    \hline
    Cam. Num. & 5 & 8 & 11 & 15 & 20 & 30 & 45 & 60 & 75 & 90 & 120 & 150 & 180 
    \\\hline
    $E [\times 10^3]$ & 4 & 3.5 & 2.5 & 2.5 & 2.5 & 2.2 & 2 & 2 & 2 & 2 & 2 & 2 & 2 
    \\\hline
    $l_r^*  [×10^{-2}]$ & 10 & 8 & 5 & 5 & 5 & 5 & 4 & 2 & 2 & 2 & 1 & 1 & 1 
    \\\hline
    $\alpha$ & 0.5 & 0.5 & 0.5 & 0.5 & 0.5 & 0.5 & 0.5 & 0.5 & 0.5 & 0.5 & 0.1 & 0.1 & 0.1
    \\\hline
    $E_d [\times 10^3]$ & 0.65 & 0.65 & 0.5 & 0.5 & 0.5 & 0.4 & 0.4 & 0.4 & 0.4 & 0.3 & 1.2 & 1.2 & 1.2
    \\\hline
    \end{tabular}
    \caption{\label{Tab:Hyperparameters} Learning rate schedule in NeDF for cases with different numbers of cameras.}
\end{table}

It is observed in Table~\ref{Tab:Hyperparameters} that for cases with a small amount of training data, more training epochs and a higher initial learning rate yield a better predicted accuracy, which is reasonable since fewer optimization steps are performed in each epoch with fewer BOS views. Therefore, more epochs are needed to achieve convergence. Additionally, a higher initial learning rate is preferred to avoid the local minimal during the course of training. 

% Additionally, the NeDF with limited training data  generally converges to a local optimum or overfit, requiring a higher learning rate to allow the optimization algorithm to search in a larger parameter space.

\section{Study on the architectures of NeDF}

In this section, we perform a comprehensive study on the effect of the NN architectures, i.e., depth and width, on the predicted accuracy as well as the robustness of NeDF. 
In all the test cases, the learning rate schedule present in Sec. \ref{sec:lr} is employed. 

% Specifically, the case with a circular camera array with 11 views is employed as representative. We first vary the depth of NeDFs from 3 to 6 while keeping the width fixed for each hidden layer, as illustrated in Table \ref{Tab:varyingLayer}. Note that the width is inherited from {\color{red}\cite{XXX}}.  The predicted accuracy represented by the metrics in Sec. \ref{sec:metrics} for this specific case is depicted in Fig. \ref{Fig:varyingLayer}.  As shown, the NeDFs with different depth are able to provide robust predictions, and achieve the best accuracy among all the tested approaches here. 

\begin{figure}[htbp]
\centering
\includegraphics[width=1\linewidth]{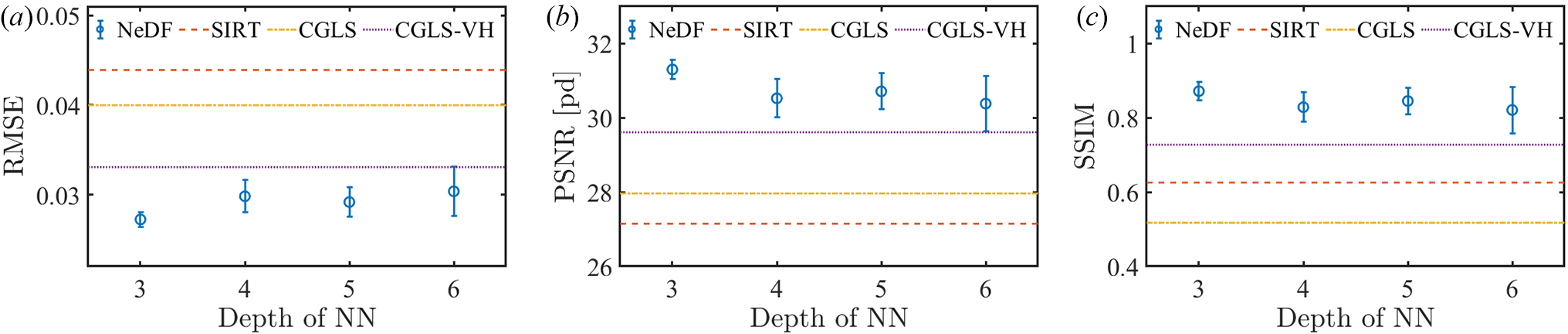}
\caption{\label{Fig:varyingLayer}Predicted accuracy for NeDFs with different depth: (\textit{a}) RMSE; (\textit{b}) PSNR; (\textit{c}) SSIM. The circle markers and errorbars represent the mean and standard deviation from 6 independent runs with different initializations of the NNs.}
\end{figure}

Specifically, the case with a circular camera array with 11 views is employed as representative. We first vary the depth of NeDFs from 3 to 6 while keeping the width fixed for each hidden layer, as illustrated in Table~\ref{Tab:varyingLayer}. Note that the width is set to be slightly larger than the dimension of the position encoding vector. The predicted accuracy represented by the metrics in Sec.~\ref{sec:metrics} for these specific cases are depicted in Fig.~\ref{Fig:varyingLayer}. As shown, the NeDFs with different depth are able to provide robust predictions, and achieve the best accuracy among all the tested approaches here.

% We verified the proposed network structure by varying the scale of the MLP including the number of layers (depth) and nodes (width). Changing the neural network scale may affect its feature extraction and generalization capabilities, and directly determines the total dimensionality of the parameters $D_\mathrm{\Theta}$ in MLP:
% \begin{equation}
%     D_\mathrm{\Theta} = \overset{\textit{First layer}}{D_{mid}(D_{in}+1)} + 
%                         \overset{\textit{Middle layer (no skip)}}{D_{mid}(D_{mid} +1)(n_{mlp}-n_s-2)} + 
%                         \overset{\textit{Middle layer (with skip)}}{D_{mid}((D_{in}+D_{mid})+1)n_s} +
%                         \overset{Output layer}{3(D_{mid}+1)}
% \label{Eq:MLPparameters}
% \end{equation}
% where $D_{in}$ is the input’s dimension. The input to the MLP of the proposed model is the positional encoding vector ($D_{in}=32$). $D_{mid}$ is the number of nodes in the middle layer. $n_{mlp}$ is the layer number of MLP. $n_s$ is the number of middle layers with skip concatenates.

% The tested configuration consisted of a circular camera array with $11$ views. Each network structure was tested independently at least six times. The results are as follows:

% \textbf{a) Neural networks with different numbers of layers}

% Table~\ref{Tab:varyingLayer} lists the network parameters for depths ranging from $3$ to $6$, and Figure~\ref{Fig:varyingLayer} shows the test results for these network structures.

\begin{table}[htbp]
    \centering
    \begin{tabular}{c|c|c|c}
    \hline
    \textbf{Depth} [layers] & \textbf{Width} [nodes] & \textbf{Skip to} [layer] & $D_\mathrm{\Theta}$ 
    \\\hline
    3 & 44 & None & 3567
    \\\hline
    4 & 44 & $3^{rd}$ & 6955
    \\\hline
    5 & 44 & $4^{th}$ & 8935
    \\\hline
    6 & 44 & $5^{th}$ & 10915
    \\\hline
    \end{tabular}
    \caption{\label{Tab:varyingLayer} NeDF with different depth and the total parameters ($D_{\Theta}$) to be optimized.}
\end{table}

% \textbf{b) Neural networks with different numbers of nodes}

\begin{figure}
\centering
\includegraphics[width=1\linewidth]{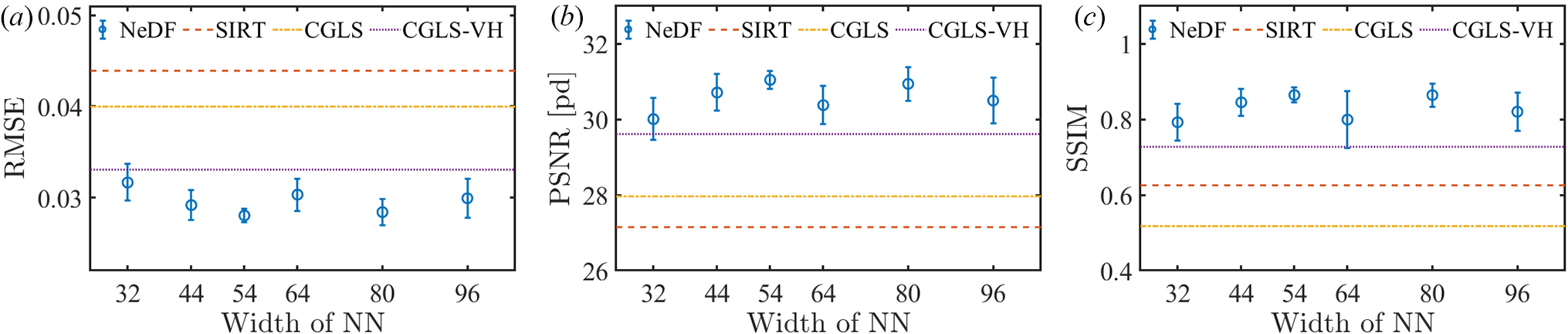}
\caption{\label{Fig:varyingNode} Predicted accuracy of NeDFs with different width: (\textit{a}) RMSE; (\textit{b}) PSNR; (\textit{c}) SSIM. The circle markers and errorbars represent the mean and standard deviation from 6 independent runs with different initializations of NNs.}
\end{figure}

It is observed in Fig. \ref{Fig:varyingLayer} that the NeDF with the depth 5 is able to provide both accurate and robust predictions. We then vary the width of NeDF with fixing the depth as 5 to further study the effect of the width on the  predicted accuracy. As shown in Table~\ref{Tab:varyingNode}, the width ranges from 32 to 96 here. In addition, the training data are the same as in the aforementioned cases.  As illustrated in Fig.~\ref{Fig:varyingNode}, the NeDFs with different width are also able to obtain robust as well as accurate predictions, which is similar as in the above cases and will not be discussed in detail for simplicity. 

% Table~\ref{Tab:varyingNode} lists the network parameters for nodes ranging from 32 to 96, and Figure~\ref{Fig:varyingNode} shows the test results for these network structures.

\begin{table}
    \centering
    \begin{tabular}{c|c|c|c}
    \hline
    \textbf{Depth} [layers] & \textbf{Width} [nodes] & \textbf{Skip to} [layer] & $D_\mathrm{\Theta}$ 
    \\\hline
    5 & 32 & $4^{th}$ & 5347
    \\\hline
    5 & 44 & $4^{th}$ & 8935
    \\\hline
    5 & 54 & $4^{th}$ & 12585
    \\\hline
    5 & 64 & $4^{th}$ & 16835
    \\\hline
    5 & 80 & $4^{th}$ & 24883
    \\\hline
    5 & 96 & $4^{th}$ & 34467
    \\\hline
    \end{tabular}
    \caption{\label{Tab:varyingNode} NeDFs with different width and the total parameters ($\Theta$) to be optimized.}
\end{table}

% For the MLPs, an oversized network is prone to overfitting and increases computational cost, while an undersized network may fail to capture all the features of the flow field. Therefore, in this study, we moderately chose a combination of 5 layers and 44 nodes for demonstration. Figures~\ref{Fig:varyingLayer} and \ref{Fig:varyingNode} show that the reconstruction results from this network structure, across multiple repetitions, consistently outperformed traditional algorithms, with errors within a small range, thus making the chosen structure parameters and results representative. 
We would like to discuss here that the fluctuations in errorbars in each test case are primarily due to the fact that sparse-view reconstruction is inherently an underdetermined problem. Further, the NeDFs with different architectures (depth and width) outperform all the traditional methods employed in this specific case, demonstrating the accuracy as well as robustness of the present model. Finally, we note that 
optimizing the NeDF architectures for all the test cases to achieve better accuracy is beyond the scope of the current study. We will leave this interesting topic as future studies. 

% However, the impact of network structure is not the focus of this study, and thus we will not explore it further, although it can be a direction for future research.

% \bibliographystyle{alpha}
% \bibliography{sample}